\documentclass[prx,onecolumn,nofootinbib,citeautoscript,eqsecnum,10pt,longbibliography,notitlepage]{revtex4-2}

\usepackage{graphicx}
\usepackage{dcolumn}
\usepackage{bm}
\usepackage{color} 

\usepackage[papersize={8.5in,11in}]{geometry}

\usepackage{color}
\definecolor{darkblue}{rgb}{0.,0.,0.4}
\definecolor{darkred}{rgb}{0.5,0.,0.}
\definecolor{BlueViolet}{RGB}{138,43,226}
\definecolor{SkyBlue}{RGB}{30,144,255}
\definecolor{DarkGreen}{RGB}{0,100,0}
\usepackage[pdftex,colorlinks=true,linkcolor=darkblue,citecolor=blue,urlcolor=darkred]{hyperref}

\usepackage{physics}
\usepackage{cancel}
\usepackage[version=4]{mhchem}

\def \nn{\nonumber \\}

\begin{document}

\title{Zero sound and plasmon modes for non-Fermi liquids}

\author{Ipsita Mandal}
\affiliation{Institute of Nuclear Physics, Polish Academy of Sciences, 31-342 Krak\'{o}w, Poland}

\begin{abstract}
We derive the quantum Boltzmann equation (QBE) by using generalized Landau-interaction parameters, obtained through the nonequilibrium Green's function technique. This is a generalization of the usual QBE formalism to non-Fermi liquid (NFL) systems, which do not have well-defined quasiparticles. We apply this framework to a controlled low-energy effective field theory for the Ising-nematic quantum critical point, in order to find the collective excitations of the critical Fermi surface in the collisionless regime. We also compute the nature of the dispersion after the addition of weak Coulomb interactions. The zero angular momentum longitudinal vibrations of the Fermi surface show a linear-in-wavenumber dispersion, which corresponds to the zero sound of Landau's Fermi liquid theory. The Coulomb interaction modifies it to a plasmon mode in the long-wavelength limit, which disperses as the square-root of the wavenumber. Remarkably, our results show that the zero sound and plasmon modes exhibit the same behaviour as in a Fermi liquid, although an NFL is fundamentally different from the former.
\end{abstract}
\maketitle

\tableofcontents

\section{Introduction}

An intriguing emergent phenomenon in strongly-coupled electron-systems is the existence of non-Fermi liquids (NFLs), which are metallic states lying beyond the framework of Landau's Fermi liquid theory. Well-known scenarios where the NFLs arise are when finite-density fermions interact with a critical boson arising at a quantum critical point \cite{metlsach1,metlsach,Lee-Dalid,ips-uv-ir1,ips-uv-ir2,ips-fflo,ips-rafael}, or with massless gauge fields \cite{olav,ips2,ips3,ips-nfl-u1}. NFLs can also arise at band-touching points of semimetals, due to the effect of long-range Coulomb interactions \cite{Abrikosov,moon-xu,rahul-sid,ips-rahul,malcolm-bitan,ips-birefringent}.
The NFL character manifests itself in various ways like transport properties
\cite{ips-subir,ips-c2,ips-hermann,ips-hermann2,ips-hermann3}, and a changed susceptibility towards superconducting instability \cite{ips2,ips-sc,ips-c2}
compared to the Fermi liquid case.

The Ising-nematic order is associated with electronic correlations which spontaneously break the square lattice symmetry to that of a rectangular lattice \cite{metlsach1,sachdev_2011}. In other words, it describes a Pomeranchuk transition where the four-fold rotational symmetry of the Fermi surface (i.e., the symmetry for rotations by $\pi/2$) is broken down to two-fold rotations, such that the x- and y-directions become anisotropic. This broken symmetry is associated with an Ising order parameter, which can be represented by a real scalar boson $\phi$, centred at wavevector $\mathbf{Q} = 0$. The results from a number of experiments \cite{yoichi,hinkov,Kohsaka,Daou} are believed to indicate the presence of this ordering in the normal state of the cuprate superconductors, which makes this critical point particularly important for analytical studies. In addition, the NFL behaviour originating from the quantum critical point, does not allow us to describe the system theoretically using the usual Landau's Fermi liquid description in terms of quasiparticles. One needs to devise controlled approximations to figure out the universal scalings of this NFL phase \cite{Lee-Dalid,ips-uv-ir1,ips-uv-ir2}.

In this paper, we will find the collective excitations of this NFL in the collisionless regime. The system has a well-defined Fermi surface, but no well-defined quasisparticles \cite{Lee-Dalid,ips-uv-ir1,ips-uv-ir2,ips-subir}. In particular, our aim is to identify the analogue of the zero sound (defined for interacting Fermi liquids), corresponding to the natural oscillations of the Fermi surface in the zero angular momentum channel, resulting from interactions. We will also investigate the effect of weak Coulomb interactions on the collective modes, which give rise to plasmons in normal metals.

In order to compute the dispersions of collective modes in a Fermi liquid, the usual quatum Boltzmann equation (QBE) formalism in terms of the Fermi distribution function works, which hinges on the existence of well-defined quasiparticles. However, since quasiparticles are destroyed in an NFL, this framework cannot be applied to find the collective excitations for NFLs with a critical Fermi surface. To overcome this difficulty, we will use the nonequilibrium Green's function technique to derive a generalized QBE, introduced by Ref.~\cite{prange,kim_qbe}. This formalism uses a generalized Landau-interaction, which has a frequency-dependence, in addition to the usual angular-dependence, due to the retarded nature of the gapless bosonic interactions. However, the earlier works used a random-phase approximation (RPA)
by employing a large-$N$ expansion (i.e., by introducing $N$ favours of fermions with $N \rightarrow \infty $), and subsequently it was proved that the infinite-flavour limit is not described by a mean-field theory (due to the large residual quantum fluctuations of the Fermi surface \cite{SSLee}).
This necessitated the development of a controlled approximation incorporating the dimensional regularization scheme, using a patch coordinate on the Fermi surface \cite{Lee-Dalid,ips-uv-ir1}. We will use the one-loop corrected bosonic propagator obtained from this controlled framework, and derive the QBE. Finally, we will capture the effect of weak Coulomb interactions, and investigate the behaviour of plasmons, by adding appropriate terms to the QBE.

\section{Keldysh formalism for the Ising-nematic critical point}

We consider the quantum phase transition brought about by the Ising-nematic order parameter. The symmetry breaking is characterized by a real scalar field $\phi$, which
is described by the Ginzburg-Landau Lagrangian action \cite{metlsach1,sachdev_2011}
\begin{align}
\label{ac1}
S_\phi & =  \frac{1}{2} \int d\tau \,dx \, dy\left[  \phi
\left(- \partial_\tau ^2 -  c_\phi ^2\partial_x ^2 -  c_\phi ^2 \partial_y ^2
\right) \phi
 + r_\phi \,\phi^2 
+\frac{u_\phi \,\phi^ 4 } {12}
\right],
\end{align}
in the position space and in imaginary time $\tau$
\footnote{Here we use the convention that $-i \grad_{\mathbf r} $ is the momentum operator in position space, such that the Fourier transform of a field $\Phi$ is defined as $\Phi(\tau,\mathbf r) =
\int \frac{dk_0\,dk_x\,dk_y}  {(2\pi)^{3/2}}
\, \Phi(k_0,\mathbf k) \,e^{ -i\,k_0 \tau + i\,\mathbf k \cdot \mathbf r}$.},
where $c_\phi$ is the bosonic velocity, $ r_\phi$ is the tuning parameter across the phase transition, and $u_\phi$ is the coupling constant for the quartic term.
All couplings, apart from $r_\phi$, can be scaled away or set equal to unity. 
The quantum phase transition occurs at $ r_\phi =0$ at temperature $T=0$, and this transition is in the same universality class as the classical two-dimensional Ising model. However, coupling this order parameter boson to gapless fermions changes the nature of the quantum critical fluctuations, such that the quantum critical point (QCP) in the fermion-boson system is no longer in the usual Ising universality class.
These quantum effects play a crucial role not only at the QCP, but also in the fan-shaped quantum critical region emanating from the QCP.

\begin{figure}
\centering
\includegraphics[width = 0.25 \textwidth]{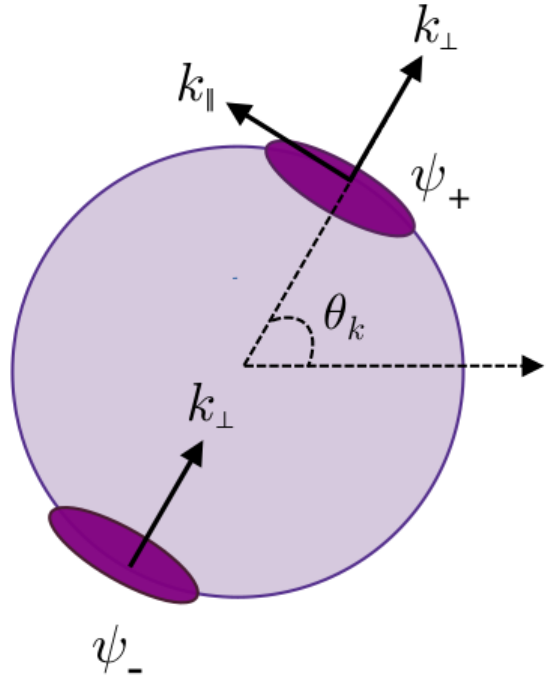} 
\caption{\label{figpatch}
$\psi_{+}$ denotes the fermion field located at the upper purple patch, centred at an angle $\theta=\theta_k$, with respect to the $x$-axis of the global coordinate system for a circular Fermi surface (denoted by the purple ring). $\psi_{-}$ denotes the fermionic field in the lower purple patch, centred at the antipodal point with $\theta= \pi+\theta_k$, whose tangential momentum is parallel to that at $\theta_k$, and needs to be included in the effective action. Although we show here the patch construction for a circular Fermi surface for the sake of simplicity, this can be applied to any Fermi surface of a generic shape, as long as it is locally convex at each point.
}
\end{figure}

Now let us focus on writing the effective action for electrons in a convenient coordinate system. We will use the patch theory used in Ref.~\cite{metlsach1,sachdev_2011,Lee-Dalid,ips-uv-ir1,ips-uv-ir2,ips-nfl-u1}, which hinges on the fact that fermions in one region of the
momentum space near the Fermi surface are primarily coupled
with a critical boson, whose momentum $\mathbf q$ is tangential to the Fermi
surface (i.e., along $k_\parallel$ as shown in Fig.~\ref{figpatch}). In other words, fermions in different momentum patches (except for
the ones at the antipodal points) are decoupled from each other in the low-energy limit. Consequently, one can extract observables that are local in momentum space (e.g. Green’s functions) from local patches in momentum space, without having to refer to the global properties of the Fermi surface.
Hence, the action for the fermions is captured by
\begin{align}
\label{ac2}
S_f &= \sum \limits_{s=\pm}
\int_{k} \psi^\dagger_{s}(k) 
\left [ -i \,k_0 + s\,v_F \,k_{\perp} +  
\frac{k_{\parallel}^2} {2\,m} \right ] \psi_{s}(k) \,,
\end{align}
where the fermionic fields $\psi_{\pm}$ denote the (right-)left-moving fermions on the antipodal patches (see Fig.~\ref{figpatch}). We have used the shorthand notations $k \equiv (k_0, \mathbf k)$ ($k_0 $ is the Matsubara frequency)
and $ \int_k \equiv \int \frac{dk_0\,dk_{\parallel}\,dk_\perp } {(2\,\pi)^3}$.
Here, the patch coordinate involves expanding the fermion momentum about the local Fermi momentum $k_F$, such that $ k_{\perp}$ is directed perpendicular to the Fermi surface, and $k_{\parallel}$ is tangential to it.

In the next step, we need to couple the Ising-nematic order parameter boson to the fermions. A simple convenient choice, deduced through symmetry considerations, is captured by the action \cite{sachdev_2011}
\begin{align}
\label{ac3}
 S_{int} 
= \tilde e
\int_{k, q} \left(  \cos k_x -\cos k_y  \right)
\phi(q) \,\psi^\dagger (k+q) \, \psi(k) \,,
\end{align}
where $\tilde e$ is the fermion-boson coupling strength, and this action is written in the global coordinates (rather than the patch coordinates).
Here, $\left(  \cos k_x -\cos k_y  \right)$ is the nematic d-wave form-factor, reflecting the fact that the nematic order parameter couples to the quadrupolar distortion of the Fermi surface. The integral over $q$ is over small momenta, while that over $k$ extends over the entire Brillouin zone.
Now, using the patch coordinates, and keeping only the leading order term, we get
$\left(  \cos k_x -\cos k_y  \right) \simeq \cos k_F$. Redefining the coupling constant as $e = \tilde e \, \cos k_F$, we can then rewrite the fermion-boson interaction in the patch coordinates as
\begin{align}
 \tilde S_{int} 
= e\,  \sum \limits_{s=\pm}
\int_{k, q} \phi(q) \,\psi^\dagger_{s}(k+q) \, \psi_{s}(k) \,.
\end{align}

After an appropriate rescaling of the energy and momenta, and dropping irrelevant terms \cite{Lee-Dalid,ips-uv-ir1} (considering $S_f$ to determine the engineering dimensions of the various terms) like $\left (k_0^2 + k_\perp^2 \right) \phi(-k) \,\phi(k)$ and $\phi^4$,
the effective action for the Ising-nematic critical point in the patch construction 
reduces to
\begin{align}
S_{tot} & = S_f + \tilde S_\phi + \tilde S_{int} \,,\quad
\tilde S_\phi  = \frac{1}{2} \int_{k}\,k_{\parallel}^2 \,\phi(-k) \,\phi(k) \,.
\end{align}
We rewrite the above action in terms of the two-component spinor $\Psi$,
where
\begin{align}
\Psi ^T(k) = \left( 
\psi_{+}(k)\quad  \quad \psi_{-}^\dagger(-k)\right), \quad \bar \Psi (k) 
= \Psi^\dagger (k)\,\sigma_2 \,, 
\end{align}
such that
\begin{align}
S_f &=  
\int_{k} \bar \Psi (k) 
\,i \left (  -\sigma_2  \,k_0 +  \sigma_1\, \delta_k \right ) 
  \Psi(k) \,, \quad
S_{int} 
= i\,e \int_{k, q} \phi(q)\,
\bar \Psi (k+q)  \,\sigma_1 \,  \Psi (k) \,,\quad
\delta_k = v_F\,k_{\perp} +  \frac{k_{\parallel}^2} { 2\, m} \,.
\end{align}
From the above action, we define the bare Matsubara Green’s function for the fermions as
\cite{kamenev}
\begin{align}
\mathcal G(i\,k_0,\mathbf k)
= -\left \langle \Psi(k) \,\bar \Psi(k) \right \rangle ,
\end{align}
whose bare value is given by
\begin{align}
\label{eqbareGF}
{\mathcal G}^{(0)}(i\,k_0,\mathbf k)
= \left[ i \left (  \sigma_2  \,k_0 -  \sigma_1\, \delta_k \right ) \right ]^{-1}
= -i\,\frac{ \sigma_2 \,k_0 - \sigma_1 \,\delta_k}
{k_0^2 +\delta_k^2} \,.
\end{align}
We note that this is the Fourier transform of the imaginary time Matsubara Green's function, given by
\begin{align}
& \mathcal G_{\mu \nu}(\tau,\tau',\mathbf k)  
\equiv \mathcal G_{\mu \nu}(\tau -\tau',\mathbf k)
= - \left \langle T_\tau\, \Psi_\mu (\tau,\mathbf k) \,
\bar \Psi_{\nu}(\tau',\mathbf k) \right \rangle \nn
& = -  
\theta (\tau -\tau') \left \langle \Psi_\mu(\tau,\mathbf k) \,
\bar \Psi_\nu(\tau',\mathbf k) \right \rangle 
+
\theta (\tau' -\tau) \left \langle \bar  \Psi _\nu(\tau',\mathbf k) \,
 \Psi_\mu(\tau,\mathbf k) \right \rangle ,
\end{align}
where $\mu, \nu \in \lbrace 1,2 \rbrace $, and $T_\tau$ is the time-ordering operator. More explicitly, we have the relation
$\mathcal G_{\mu \nu}(i\,k_0,\mathbf k) =\int \frac{dk_0} {\sqrt{2\pi}}\, 
e^{i\,k_0\left (\tau-\tau'\right)} 
\,\mathcal G_{\mu \nu}(\tau -\tau',\mathbf k)$.

The dressed boson propagator, which
includes the one-loop self-energy $$\Pi_1(q)
\equiv  (i\,e)^2\,
\int_{k} 
\text{Tr}
\left [ {\mathcal{G}}^{(0)}(k + q)\,\sigma_1\, {\mathcal{G}}^{(0)}(k)\,\sigma_1 \right ]
= \frac{ e^2 \,m\,|q_0| } { \pi\,v_F\,| q_\parallel |},$$ is given by
\cite{Lee-Dalid,ips-uv-ir1}
\begin{align}
\label{eqbareD}
\mathcal{D}(i\,q_0,\mathbf{q}) &
= -\left \langle \phi(q) \, \phi(-q) \right \rangle 
=  \left[D_0^{-1}(q) -\Pi_1(q) \right ]^{-1}
= - \frac{1}
{ q_{\parallel}^2 +\frac{ e^2 \,m\,|q_0| } { \pi\,v_F\,| q_\parallel |}}\,,
\end{align}
because the zeroth order propagator is $D_0^{-1}(q) = -\,q_\parallel^2$.
Using this form, the one-loop fermion self-energy in the Matsubara space is found to be
\begin{align}
\label{eqfse}
\Sigma(k) = 
- \frac{i \,e^{4/3} \,\text{sgn}(k_0)\,|k_0|^{2/3} 
\, \sigma_2} 
{ 2 \,\sqrt{3}\, \pi ^{2/3} \,k_F^{1/3} } \,,
\end{align}
such that the one-loop corrected Green's function (using the Landau-damped boson propagator) is given by
$ \mathcal{G}^{-1} (k) = \left[ {\mathcal G}^{(0)}(k) \right]^{-1} - \Sigma (k)$.

To deal with a nonequilibrium situation, we need to formulate the action on the closed-time Keldysh contour \cite{kamenev}, where we evolve the system according to the Heisenberg representation for operators via the following sequence: start from the distant past, evolve forward to the time of physical
interest, and then evolve backward to the distant past. In order to rewrite the action in real time $t$ (where $\tau = i\, t$), an action $\mathcal S_{M}(\tau)$ in the Matsubara space is transformed as
$ -i\,\mathcal S (t) = \mathcal S_M (\tau)  \vert_{\tau \rightarrow \,i\, t}$.
Using the superscripts ``$+$'' and ``$-$'' to refer to the fields residing on the forward and backward parts of the Keldysh contour, respectively, we obtain the action as $S_{K} = S_{K,f} + S_{K,b}+S_{K,int} ,$ where
\begin{align}
 S_{K,f}  & =  
\int_{-\infty}^{\infty}  d t \int \frac{ d^2 \mathbf{k}} {(2\pi)^2}
  \Bigg[ {\bar{\Psi}}^+ (t,\mathbf{k}) \,
 i \left(  \sigma_2 \,\partial_t - \sigma_1\,\delta_k \right)
\Psi^+ (t,\mathbf{k}) 
- {\bar{\Psi}}^{-}(t,\mathbf{k}) \,
i \left( \sigma_2\, \partial_t - \sigma_1\,\delta_k \right) \Psi^{-}(t,\mathbf{k})
\Bigg],\nonumber \\
 S_{K,\phi}  & =  \frac{1}{2} \int_{-\infty}^{\infty} 
dt \int \frac{ d^2 \mathbf{k}} {(2\pi)^2}
\Bigg[
\phi^{+}(t,-\mathbf{k})
\left(-\partial_t^2 - k_\parallel^2 \right)
\phi^{+}(t,\mathbf{k})
-\phi^{-}(t,-\mathbf{k})\left( -\partial_t^2 - k_\parallel^2\right)
\phi^{-}(t,\mathbf{k})\Bigg]\,,\nn
S_{K,int} &=
- i\, e \int_{-\infty}^{\infty} d t 
\int d^{2} \mathbf{r}
\Bigg[
\phi^{+}(t,\mathbf{r})\,
{\bar{\Psi}}^{+} (t,\mathbf{r})\,\sigma_1\,
\Psi^{+}(t,\mathbf{r})-  \phi^{-}(t,\mathbf{r})\,
{\bar{\Psi}}^{-}(t,\mathbf{r})\,\sigma_1\,\Psi^{-}(t,\mathbf{r}) \Bigg]\,.
\label{plusminus}
\end{align}
We note that the relative minus signs come from reversing the direction of the time integration on the backward part of the contour. The nonequilibrium formalism automatically includes several different Green’s functions, depending on the location of each time argument on the contour. The lesser, greater, time-ordered, and anti-time-ordered Green’s functions are defined by the expressions
\begin{align}
\label{eq4gf}
 G_{\mu \nu}^< (t_1,\mathbf r_1; t_2,\mathbf r_2) &= -i
\left \langle
\Psi^+ _\mu (t_1,\mathbf r_1)\,
 {\bar \Psi}^-_\nu (t_2,\mathbf r_2) \right \rangle, \quad
G_{\mu \nu}^> (t_1,\mathbf r_1; t_2,\mathbf r_2) = -i
\left \langle
\Psi^-_\mu (t_1,\mathbf r_1)\,{\bar \Psi}^+_\nu (t_2,\mathbf r_2) \right \rangle,\nn
 G_{\mu \nu}^{\mathcal{T}} (t_1,\mathbf r_1; t_2,\mathbf r_2) &= -i
\left \langle
 \Psi^+ _\mu (t_1,\mathbf r_1)\,{\bar \Psi}^+_\nu (t_2,\mathbf r_2) \right \rangle
= \theta(t_1-t_2)\,G_{\mu \nu}^> (t_1,\mathbf r_1; t_2,\mathbf r_2)
+ \theta(t_2-t_1)\,G_{\mu \nu}^< (t_1,\mathbf r_1; t_2,\mathbf r_2),
\nn \text{and }
G_{\mu \nu}^{ \bar{\mathcal{T}}} (t_1,\mathbf r_1; t_2,\mathbf r_2) 
&= -i
\left \langle \Psi^- _\mu (t_1,\mathbf r_1)\,{\bar \Psi}^-_\nu (t_2,\mathbf r_2) \right \rangle
= \theta(t_2-t_1)\,G_{\mu \nu}^> (t_1,\mathbf r_1; t_2,\mathbf r_2)
+ \theta(t_1-t_2)\,G_{\mu \nu}^< (t_1,\mathbf r_1; t_2,\mathbf r_2),
\end{align}
respectively.
Similarly, for the bosons, we define
\begin{align}
D^< (t_1,\mathbf r_1; t_2,\mathbf r_2) &= -i
\left \langle
\phi^+ (t_1,\mathbf r_1)\, {\phi}^- (t_2,\mathbf r_2) \right \rangle, \quad
D^> (t_1,\mathbf r_1; t_2,\mathbf r_2) = -i
\left \langle
\phi^- (t_1,\mathbf r_1)\,{\phi}^+ (t_2,\mathbf r_2) \right \rangle,\nn
 D^{\mathcal{T}} (t_1,\mathbf r_1; t_2,\mathbf r_2) &= -i
\left \langle
 \phi^+  (t_1,\mathbf r_1)\,{\phi}^+ (t_2,\mathbf r_2) \right \rangle
= \theta(t_1-t_2)\,D^> (t_1,\mathbf r_1; t_2,\mathbf r_2)
+ \theta(t_2-t_1)\,D^< (t_1,\mathbf r_1; t_2,\mathbf r_2),
\nn D^{ \bar{\mathcal{T}}} (t_1,\mathbf r_1; t_2,\mathbf r_2) 
&= -i
\left \langle \phi^-  (t_1,\mathbf r_1)\,{\phi}^- (t_2,\mathbf r_2) \right \rangle
= \theta(t_2-t_1)\,D^> (t_1,\mathbf r_1; t_2,\mathbf r_2)
+ \theta(t_1-t_2)\,D^< (t_1,\mathbf r_1; t_2,\mathbf r_2),
\end{align}

From the above Green’s functions, we can construct the so-called retarded and advanced Green’s functions
(denoted by the supercripts $R$ and $A$, respectively) via
\begin{align}
\label{eqdefRA}
G_{\mu \nu}^R(t_1,\mathbf r_1; t_2,\mathbf r_2) &= 
G_{\mu \nu}^{\mathcal{T}} (t_1,\mathbf r_1; t_2,\mathbf r_2) 
- G_{\mu \nu}^< (t_1,\mathbf r_1; t_2,\mathbf r_2)
= G_{\mu \nu}^> (t_1,\mathbf r_1; t_2,\mathbf r_2)
- G_{\mu \nu}^{\bar {\mathcal{T}}} (t_1,\mathbf r_1; t_2,\mathbf r_2),  \nn
G_{\mu \nu}^A (t_1,\mathbf r_1; t_2,\mathbf r_2) &= 
G_{\mu \nu}^{\mathcal{T}} (t_1,\mathbf r_1; t_2,\mathbf r_2) 
- G_{\mu \nu}^> (t_1,\mathbf r_1; t_2,\mathbf r_2)
= G_{\mu \nu}^< (t_1,\mathbf r_1; t_2,\mathbf r_2)
- G_{\mu \nu}^{\bar {\mathcal{T}}} (t_1,\mathbf r_1; t_2,\mathbf r_2).
\end{align}
Using these definitions, we immediately observe the relations
\begin{align}
G^R_{\mu \nu}(t_1,\mathbf r_1; t_2,\mathbf r_2)
&=\theta(t_1-t_2) \left[ 
G^>_{\mu \nu}(t_1,\mathbf r_1; t_2,\mathbf r_2)-G^<_{\mu \nu}(t_1,\mathbf r_1; t_2,\mathbf r_2)
\right],\nn
G^A_{\mu \nu}(t_1,\mathbf r_1; t_2,\mathbf r_2)
&= -\theta(t_2-t_1) \left[ 
G^>_{\mu \nu}(t_1,\mathbf r_1; t_2,\mathbf r_2)-G^<_{\mu \nu}(t_1,\mathbf r_1; t_2,\mathbf r_2)
\right] ,\nn
G^R_{\mu \nu}(t_1,\mathbf r_1; t_2,\mathbf r_2)& - G^A_{\mu \nu}(t_1,\mathbf r_1; t_2,\mathbf r_2)
= G^>_{\mu \nu}(t_1,\mathbf r_1; t_2,\mathbf r_2)-G^<_{\mu \nu}(t_1,\mathbf r_1; t_2,\mathbf r_2)\,.
\end{align}
One can write analogous equations and identities for the bosons as well.

In equilibrium, using Eq.~\eqref{eqbareGF}, the explicit expressions for the bare retarded and advanced Green's functions are given by
\begin{align}
G_{bare}^{R}( \omega_k ,\mathbf{k}) &= 
{\mathcal G}^{(0)}(i\,k_0,\mathbf k) \Big \vert_  {i\,k_0 \rightarrow \omega_k +i\,0^+}
=
 \frac{ \sigma_2 \,\omega - i\,\sigma_1 \,\delta_k}
{\left( \omega + i\, 0^{+}\right)^2 - \delta_k^2}
\,,\nn
G_{bare}^{A }(\omega_k ,\mathbf{k}) 
 & = {\mathcal G}^{(0)}(i\,k_0,\mathbf k) \Big \vert_  {i\,k_0 \rightarrow \omega_k - i\,0^+} 
 =  \frac{ \sigma_2 \,\omega - i\,\sigma_1 \,\delta_k}
{\left( \omega - i\, 0^{+}\right)^2 - \delta_k^2} \,.
\end{align}

Using Eq.~\eqref{eqbareD}, the one-loop corrected retarded and advanced bosonic Green's functions are
\begin{align}
D_{1-loop}^{R}(\omega_q,\mathbf{q}) 
& =\mathcal D(i\,q_0,\mathbf q) \Big \vert_  {i\,q_0 \rightarrow \omega_q +i\,0^+}
= -\frac{1}
{ q_{\parallel}^2
- i\,\frac{ e^2 \,m\,\omega_q } { \pi\,v_F\,| q_\parallel |}}
\nn
\text{ and } D_{1-loop}^{A}(\omega_q,\mathbf{q}) 
& =\mathcal D(i\,q_0,\mathbf q) \Big \vert_  {i\,q_0 \rightarrow \omega_q - i\,0^+}
= -\frac{1}
{ q_{\parallel}^2
+ i\,\frac{ e^2 \,m\,\omega_q } { \pi\,v_F\,| q_\parallel |}}\,,
\end{align}
respectively. While making the analytic continuation $i\,q_0 \rightarrow \omega_q +i\,0^+$,
we have used the relation
$$\text{sgn}(q_0) \equiv \text{sgn}\Big(\text{Im}(i\,q_0) \Big)
\rightarrow \text{sgn}\Big(\text{Im} \big (\omega_q+ i \,0^+ \big ) \Big)
=\text{sgn} \big ( 0^+ \big ) =1. $$
An analogous relation has been used for the case $i\,q_0 \rightarrow \omega_q - i\,0^+$ for $D_{1-loop}^{A}$.

In equilibrium, all Green's functions of a translationally invariant system (including translations both in time and space) depend only on the differences $t_{rel} =t_1-t_2$ and $\mathbf{r}_{rel} =\mathbf r_1-\mathbf r_2$, which allows us to use the Fourier space descriptions. 
The retarded and advanced fermion self-energy functions at one-loop order, after analytic continuation of Eq.~\eqref{eqfse} to real frequencies, are captured by
\begin{align}
\Sigma^{R}(\omega_k)
= - \frac{e^{4/3} \left[ \sqrt 3\,\text{sgn}(\omega_k) + i \right]
|\omega_k |^{2/3} \, \sigma_2} 
{ 4\, v_F\,\pi ^{2/3} \left( m/v_F\right)^{1/3} } \text{  and  }
\Sigma^{A}(\omega_k)
= - \frac{e^{4/3} \left[ \sqrt 3\,\text{sgn}(\omega_k) - i \right]
|\omega_k |^{2/3} \, \sigma_2} 
{ 4\, v_F\,\pi ^{2/3} \left( m/v_F\right)^{1/3} }\,,
\label{sigmaeq}
\end{align}
respectively.
Hence, the one-loop corrected Green's functions take the forms
\begin{align}
\left[ G^{R/A}(\omega_k, \mathbf{k}) \right]^{-1}
= \left[ G_{bare}^{R/A}( \omega_k ,\mathbf{k})  \right]^{-1} -\Sigma^{R/A}(\omega_k)\,.
\end{align}
Using these expressions,
the fermionic spectral function is given by
\begin{align}
A(\omega_k,\mathbf{k})&  = 
 i   \left[ G^R(\omega_k,\mathbf{k},)-G^A(\omega_k,\mathbf{k}) \right] \sigma_2
= i \left[ G^>(\omega_k,\mathbf{k},)-G^<(\omega_k,\mathbf{k}) \right] \sigma_2 .
\end{align} 
This also implies that
\begin{align}
\label{eqspectre}
-i\,G^{<}(\omega_k,\mathbf{k}) \,\sigma_2
 =  A(\omega_k,\mathbf{k})\,f_0(\omega_k )\,,\quad
i\,G^{>}(\omega_k ,\mathbf{k}) \,\sigma_2
= A(\omega_k,\mathbf{k})  \left[ 1 - f_0(\omega_k)\right ], 
\end{align}
because
\begin{align}
G^{R}(\omega,\mathbf{k}) - G^{A}(\omega,\mathbf{k})
= G^{>}(\omega,\mathbf{k}) - G^{<}(\omega,\mathbf{k})
= G^{>}(\omega,\mathbf{k}) \left(1 +  e^{-\beta\,\omega} \right),
\end{align}
where $f_0(\omega) = \frac{1} {1 + \, e^{\beta \omega}}$ is the equilibrium Fermi distribution function.
Since the diagonal components of this matrix represent the densities of states of the $\psi_+$ and $\psi_-$ fermions at a given energy $\omega_k$, the occupation number of the two types of fermions are given by
\begin{align}
n_+ (\mathbf k) & 
=\left \langle \psi_+^\dagger(t_1=0,\mathbf k)\,\psi_+(t_2=0,\mathbf k) \right \rangle
= G_{12}^{<}( t_{rel}=0 ,\mathbf{k})
=- i\left[ G^{<}( t_{rel}=0 ,\mathbf{k})\, \sigma_2  \right]_{11}
= \int \frac{d\omega_k} {2\, \pi} \,A_{11}(\omega_k,\mathbf{k})\,f_0(\omega_k )\,,\text{ and}
\nn 
n_- (\mathbf k) & 
=\left \langle \psi_-^\dagger(t_1=0,-\mathbf k)\,\psi_- (t_2=0,-\mathbf k) \right \rangle
= G_{21}^{>}( t_{rel}=0 ,-\mathbf{k})
= i\left[ G^{>}( t_{rel}=0 ,-\mathbf{k})\, \sigma_2  \right]_{22}
\nn & = \int \frac{d\omega_k} {2\, \pi} \,A_{22}(\omega_k,-\mathbf{k})
\left[ 1 - f_0(\omega_k)\right ],
\end{align}
respectively. For the bare / non-interacting system, we have $
A_{\mu \mu}(\omega_k,\mathbf{k}) 
= 2 \,\pi\, \delta \big( \omega_k +(-1)^\mu \, \delta_k  \big),$ leading to
$n_+ (\mathbf k) = f_0(\delta_k)$ and $n_- (\mathbf k) = f_0( \delta_{-k}).$

For a generic nonequilibrium situation, the Green's functions no longer enjoy temporal and spatial translation invariance, and hence must be represented as functions of both $(t_{rel}, \mathbf r_{rel})$ and
$(t=\frac{t_1+t_2} {2}, \mathbf r  = \frac{ \mathbf r_1+ \mathbf r_2} {2})$ (these coordinates are also known as the Wigner coordinates).
Let us also define $k_1\equiv (\omega_{k_1},\mathbf k_1)$ and $k_2\equiv (\omega_{k_2},\mathbf k_2)$ to be the energy-momentum variables conjugate to $(t_1,\mathbf r_1)$ and $(t_2,\mathbf r_2)$, respectively.
Additionally, let $ k =\frac{ k_1 - k_2}{2} $ and $  q=  k_1 +  k_2$ correspond to the energy-momenta conjugate to the relative coordinates $(t_{rel}, \mathbf r_{rel})$ and the centre-of-mass coordinates $(t, \mathbf r  )$, respectively. 

For a theory involving a bonafide Fermi liquid in presence of interactions, the quasiparticles are well-defined, because the imaginary part of their self-energy
$\text{Im}[\Sigma^R] \sim \omega_k^2 \ll |\omega_k| $ for small $|\omega_k|$.
This means that the equilibrium spectral function 
is sharply peaked as a function of $\omega_k$, so that ignoring the incoherent background, it can be written as
$ A(\omega_k,\mathbf{k}) 
=-2\, \text{Im}\, G^R(\omega_k,\mathbf{k}) 
= 2 \,\pi\, \delta \big( \omega_k - \xi_\mathbf{k} - \text{Re}[\Sigma^R(\omega_k,\mathbf{k})] \big)\,,$
where $ \xi_\mathbf{k}$ is the bare energy dispersion.
Using this crucial feature, if the system is not far away from equilibrium, we can construct a closed set of  equations for the fermion distribution function $ f(\omega_k,\mathbf k;t,\mathbf r)$,
which constitute the QBEs. In particular, the linearized QBEs for the fluctuation
$ \delta f(\omega_k,\mathbf k;t,\mathbf r) = f(\omega_k, \mathbf k;t,\mathbf r) - f_0(\omega_k)$, where $f_0(\omega_k)$ is the equilibrium distribution function, describe a Fermi liquid.

For the Ising-nematic quantum critical point, at one-loop order, the spectral function evaluates to
\begin{align}
A(\omega_k,\mathbf{k})
= 2\,s_2
\begin{pmatrix}
\frac{1}{\left(\omega_k +s_1 - \delta _k \right )^2+s_2^2} & 0 \\
0 & \frac{1}{\left( \omega_k +s_1 + \delta _k \right)^2+s_2^2}
\end{pmatrix},
\end{align}
where $s_1 = \frac{e^{4/3} \, \sqrt 3\,\text{sgn}(\omega_k)\,
|\omega_k |^{2/3} } 
{ 4\,v_F\, \pi ^{2/3} \left( m/v_F\right)^{1/3} }$
and $s_2 = \frac{e^{4/3} 
|\omega_k |^{2/3} } 
{ 4\,v_F\, \pi ^{2/3} \left( m/v_F\right)^{1/3} }$.
Since $|s_2|\sim |\omega_k|^{2/3} > |\omega_k|$ for small $|\omega_k|$, we find that there are no well-defined Landau quasiparticles. In other words, each diagonal component $A_{\mu \mu}(\omega_k, \mathbf{k})$ (no sum over $\mu$) of the matrix $A$ is not a peaked function of $\omega_k $ at equilibrium, unlike for Fermi liquid systems.
Due to this fact, $  \delta f(\omega_k, \mathbf k;t,\mathbf r)$ does not satisfy a closed set of equations even at equilibrium.
However, since $\Sigma^R$ is only a function of $\omega_k $,
$A_{\mu \mu}(\omega_k, \mathbf{k})$ is still a well-peaked function of $\delta_k $ around $ \delta_k = 0$ \cite{kim_qbe}. Combining this observation with the fact that $\int_{-\infty}^{\infty} \frac{d\delta_k}{2\pi} A_{\mu \mu} = 1$, 
it follows that $G^<$ and $G^>$ are sharply peaked functions of $\delta_k$. Integrating over this region of peaking, we can then define the generalized fermion distribution function $f$ (also sometimes referred to as a Wigner distribution function) as
\begin{align}
\label{eq:gdf}
\int \frac{d\delta_k}{2\pi} \,
G^<(\omega_k, {\mathbf k} ; \omega_q,\mathbf q) 
= i\,\sigma_2\,f(\omega_k, {\mathbf k} ;\omega_q, \mathbf q)\,,\quad
 \int \frac{d\delta_k }{2\pi} \,
G^>(\omega_k, {\mathbf k} ; \omega_q,\mathbf q) 
= 
i\,\sigma_2\left[ f(\omega_k,{\mathbf k};\omega_q, \mathbf q) -1\right] .
\end{align}
Using this definition of the generalized distribution function in a system without well-defined Landau quasiparticles, we now proceed to derive the QBE that governs this distribution.
Note that this method \cite{prange} differs from the usual justification of
the Landau's Fermi liquid theory, which is based on the smallness
of the decay rate (or width of the peak in $A_{\mu \mu}$). Here, the decay rate is given by $ s_2$, and hence is not small. We can use the above relations as long as the system is not far
away from the equilibrium.

\section{QBE in the collisionless regime}

Since the collective modes involve the fluctuations of the Fermi surface, we need to rewrite everything in terms of the global coordinates, rather than the patch coordinates. This means that we will now use $ k_{\perp} \rightarrow k\cos\theta_k $ and $k_\parallel \simeq k_F \,\theta_k$, where $\theta_k$ is the angle that the Fermi momentum of the patch makes with the $x$-axis. Here, we will focus on the zero temperature limit (i.e., $T=0$).

The four types of Green's functions of Eq.~\eqref{eq4gf}, and the corresponding self-energies, can be expressed as the components of the $2\times 2$ matrices as follows \cite{mahan2013many}:
\begin{align}
\tilde {\mathcal{G}} =
\begin{pmatrix}
G^{\mathcal T} & -G^< \\
G^> & -G^{\bar{\mathcal T}}\\
\end{pmatrix}, \quad
\tilde {\Sigma} =
\begin{pmatrix}
\Sigma^{\mathcal T} & - \Sigma^< \\
\Sigma^> & -\Sigma^{\bar{\mathcal T}}\\
\end{pmatrix}.
\end{align}
The matrix $\tilde {\mathcal{G}}$ obeys the Dyson's equation
\begin{align}
\tilde {\mathcal{G}}(\zeta_1,\zeta_2)
= \tilde {\mathcal{G}}_0(\zeta_1-\zeta_2)
+ \int d\zeta _3 \int d\zeta_4 \,
\tilde {\mathcal{G}}_0(\zeta_1-\zeta_3)\,\tilde {\Sigma}(\zeta_3,\zeta_4)
\, \tilde {\mathcal{G}}(\zeta_4,\zeta_2)\,,
\end{align}
where we have used the shorthand notation $ \zeta_i \equiv (t_i, \mathbf r_i) $.

Using the equations of motion for the fermionic operators, it can be shown that $\tilde {\mathcal{G}}  $ satisfies equations of motion of the form
\begin{align}
& \left [ i\, \sigma_2\,\partial_{t_1} - i\,\sigma_1\,H_0 ({\mathbf  r_1}) \right ] 
\tilde {\mathcal{G}} (\zeta_1,\zeta_2)
=\delta (\zeta_1-\zeta_2) \,\mathcal{I}
+ \int d\zeta_3\, \tilde {\Sigma}'(\zeta_1,\zeta_3)\,\tilde {\mathcal{G}} (\zeta_3,\zeta_2)\,,
\nn & 
\left [ -i\, \sigma_2\,\partial_{t_2} - i\,\sigma_1\, H_0 ({\mathbf  r_2 }) \right ] 
\tilde {\mathcal{G}} (\zeta_1,\zeta_2)
=\delta (\zeta_1-\zeta_2) \,\mathcal{I}
+ \int d\zeta_3\, \tilde {\mathcal{G}} (\zeta_1,\zeta_3)\,\tilde {\Sigma} (\zeta_3,\zeta_2)\,,
\end{align}
where $H_0 (\mathbf r_i) = \frac{-\partial^2_{\mathbf r_i} - k_F^2} {2\,m} $.
For our purpose, it is sufficient to consider the equations of motion for $G^<$ only, which are then given by
\begin{align}
\label{eqgless1}
\left [ i\,\sigma_2\, \partial_{t_1} -i\,\sigma_1\, H_0 ({\mathbf  r_1}) \right ] \left[- G^< (\zeta_1,\zeta_2) \right]
& = \int d\zeta_3\, \tilde {\Sigma}_{1 1}(\zeta_1,\zeta_3)\,\tilde {\mathcal{G}}_{12} (\zeta_3,\zeta_2)
+ \int d\zeta_3\, \tilde {\Sigma}_{1 2}(\zeta_1,\zeta_3)\,\tilde {\mathcal{G}}_{22} (\zeta_3,\zeta_2)
\nn  \Rightarrow
\left [ i\,\sigma_2\, \partial_{t_1} - i\,\sigma_1 \,H_0 ({\mathbf  r_1}) \right ]  G^< (\zeta_1,\zeta_2) 
 & = \int d\zeta_3 \, \Sigma^{\mathcal T}(\zeta_1,\zeta_3)
 \,G^< (\zeta_3,\zeta_2) 
- \int d\zeta_3 \,\Sigma^<(\zeta_1,\zeta_3) 
\,G^{\bar{\mathcal T}} (\zeta_3,\zeta_2) \,,
\end{align}
and
\begin{align}
\label{eqgless2}
\left [ -i\, \sigma_2\,\partial_{t_2} -i\,\sigma_1\, H_0 ({\mathbf  r_2}) \right ]  
G^< (\zeta_1,\zeta_2)  
 & = \int d\zeta_3\, G^{\mathcal T}(\zeta_1,\zeta_3)\,
 \,\Sigma^< (\zeta_3,\zeta_2)
- \int d\zeta_3\, G^<(\zeta_1,\zeta_3)
\,\Sigma^{\bar{\mathcal T}} (\zeta_3,\zeta_2) \,.
\end{align}

We take the difference of Eqs.~\eqref{eqgless1} and \eqref{eqgless2}, and
use the relations
\begin{align}
& G^{\mathcal T} = G^{re}  + {1 \over 2} (G^< + G^>) \ , \quad
G^{\bar {\mathcal T }} = {1 \over 2} (G^< + G^>) - G^{re} \,,
\quad G^{re} \equiv  \frac{ G^R + G^A } {2}\,,
\quad \Sigma^{re} \equiv  \frac{ \Sigma^R + \Sigma^A } {2}\,, \nn
& \Sigma^{\mathcal T} = \Sigma^{re}  + {1 \over 2} (\Sigma^< + \Sigma^>) \ , \quad
\Sigma^{\bar {\mathcal T }} = {1 \over 2} (\Sigma^< + \Sigma^>) - \Sigma^{re} \,,
\quad \Sigma^{re} \equiv  \frac{ \Sigma^R + \Sigma^A } {2}\,,
\quad \Sigma^{re} \equiv  \frac{ \Sigma^R + \Sigma^A } {2}\,,
\end{align}
implied by Eq.~\eqref{eqdefRA}.
This leads to
\begin{align}
& i \left[ \sigma_2\,\partial_{t_1} -\sigma_2 \,\partial_{t_2} 
-\sigma_1\, H_0 ({\bf r_1}) + \sigma_1 H_0 ({\bf r_2})  \right ] G^< ( \zeta_1,\zeta_2)
\nn &=
 \int d \tilde t\,d\tilde{\mathbf r} \, 
\Bigl [  \Sigma^{re} (\zeta_1,\zeta_3) \,G^< (\zeta_3,\zeta_2)
+ \Sigma^< (\zeta_1,\zeta_3) \,G^{re} (\zeta_3,\zeta_2)
- G^< (\zeta_1,\zeta_3) \, \Sigma^{re} (\zeta_3,\zeta_2) 
-G^{re}(\zeta_1,\zeta_3) 
\,\Sigma^< (\zeta_3,\zeta_2)  \nn
& \hskip 1.75cm
+ \frac{\Sigma^> (\zeta_1,\zeta_3)
\, G^< (\zeta_3,\zeta_2)
+  G^< (\zeta_1,\zeta_3) 
\,\Sigma^> (\zeta_3,\zeta_2)} {2}
- \frac{ \Sigma^< (\zeta_1,\zeta_3) 
\,G^> (\zeta_3,\zeta_2)
+ G^> (\zeta_1,\zeta_3) 
\,\Sigma^< (\zeta_3,\zeta_2) } 
{2}   \Bigr ] \, .
\end{align}

Not too far from the equilibrium, we can linearize the above assuming that
$\delta {\tilde G} = {\tilde G} - {\tilde G}_0$ and
$\delta {\tilde \Sigma} = {\tilde \Sigma} - {\tilde \Sigma}_0$
are small, where ${\tilde G}_0$ and ${\tilde \Sigma}_0$ are
the equilibrium fermionic Green's function and self-energy matrices, respectively.
Since we are focussing on perturbations around the Fermi momentum $k_F$,
we have $\mathbf {k} = \left( k_F + |{\mathbf k}|\right) \hat{{\mathbf k}} $, and $ |{\mathbf k}|,|\mathbf q | \ll k_F$, where  $\mathbf q  = 
\mathbf k_1 + \mathbf k_2$ and $ \left( k_F + |{\mathbf k}|\right) \hat{{\mathbf k}}
= \frac{\mathbf k_1 -\mathbf k_2} {2} .$ Hence, the functional dependence of $f(\omega_k, \mathbf k ;\omega_q, \mathbf q)$ on $\mathbf k$ is effectively only via the angle $\theta_{qk} = \theta_q-\theta_k$ (i.e., the angle between ${\bf k}$ and ${\bf q}$), which we symbolically express by using the notation
$f(\omega_k, \theta_{qk} ;\omega_q, \mathbf q)$.

The Fourier-transformed linearized equation for $\delta G^< (k, q)$ can now be written as
\begin{align}
& \left (  i\,\sigma_2\,\nu - i\,\sigma_1 \,v_F \,|{\mathbf q}| \cos\theta_{kq}
\right )   \delta G^< (k,q) 
- \Sigma_{0}^{re} (k + q/2)\,\delta G^< (k,q)
-  \delta (\Sigma^{re} (k,q))\,G^<_0 (k - q/2)
\nn & 
+ G^<_0 (k + q/2)\, \delta (\Sigma^{re} (k,q))
+ \delta G^< (k,q)\,\Sigma_{0}^{re}  (k - q/2) 
-  \Sigma^<_0 (k + q/2)\, \delta (G^{re} (k,q)) 
-\delta \Sigma^< (k,q)\,G^{re}_0 (k - q/2)
\nn &  +  G^{re}_0 (k + q/2)\, \delta \Sigma^< (k,q)
+  \delta (G^{re} (k,q)) \,\Sigma^<_0 (k - q/2)   = I_{coll} ,
\end{align}
where $\nu =  \omega_q $.
The shorthand notations $k$ and $q$ stand for $(\omega_k,{\mathbf k})$ and $(\omega_q,\mathbf q)$, respectively.
Furthermore, the subscript ``$0$'' denotes the equilibrium values, and $I_{coll}$
is the collision integral. In the following, we will set $I_{coll} =0$, as we are interested in the collisionless regime where the zero sound mode exists.
The collision integral in the transport equation determines a typical collision time $\tau$. If we are interested in phenomena that occur on a time scale smaller than $\tau$, i.e. for frequencies $\omega \gg 1/\tau$, then the collision integral can be safely neglected. The solution of this equation, in the absence of external fields, gives information about the collective vibrations of the Fermi surface.
We also note that $\Sigma^R_0$ is independent of momentum [cf. Eq.~\eqref{sigmaeq}].

We now perform a $\int \frac{d\delta_{k}}{2\,\pi}$ integration on either side of the above equation, and use the fact that 
$ \int \frac{d\delta_k } {2\,\pi} \,G^{re}(k,q) = 0$ \cite{prange,kim_qbe}.
Thus the sixth to ninth terms on the LHS of the QBE vanish after
the integration. On using the relations in Eq.~\eqref{eq:gdf}, we get
the LHS as
\begin{align}
&  -i \int \frac{d\delta_{k}} {2\,\pi} 
\Bigg[
i \left ( \sigma_2\,\nu -\sigma_1 \,v_F\, |\mathbf q| \cos \theta_{qk}
\right )   \delta G^< (k,q) 
- \Sigma_{0}^{re} ( \omega_k + \nu/2)\,\delta G^< (k,q)
-  \delta \Sigma^{re} (k,q)\,G^<_0 ( k - q/2)
 \nn &  \hspace{ 1.75 cm}
+ G^<_0 (k + q/2)\, \delta \Sigma^{re} (k,q)
+ \delta G^< (k,q)\,\Sigma_{0}^{re} (\omega_k - \nu/2) \Bigg]
\nn & =  \Big[
i \left ( \nu- \sigma_3 \,v_F\, |\mathbf q| \cos \theta_{qk} \right )  
-\Sigma_{0}^{re} (\omega_k + \nu/2)\,\sigma_2 
+ \sigma_2 \,\Sigma_{0}^{re} (\omega_k - \nu/2)
\, \Big ]
\delta f(\omega_k, \theta_{qk}; \nu,\mathbf q ) 
 \nn & \quad
-  \delta \Sigma^{re} (k,q)\,\sigma_2\,\Theta (-\omega_k + \nu/2)
+ \sigma_2\,\Theta (-\omega_k - \nu/2 )\, \delta \Sigma^{re} (k,q)\,,
\label{step1}
\end{align}
noting that the equilibrium value of $f$ is $f_0(k) =\Theta(-\omega_k)$ at $T=0$.

With $n_0 (\nu)= 
\frac{1} {e^{\nu/T}-1} $ representing the equilibrium boson distribution function, for small deviations from equilibrium, we have the expressions
\begin{align}
& \frac{\Sigma^< (\omega_{k'}, \mathbf k';\omega_q , \mathbf q)} 
{ e^2\,m }
\nn & 
 = -i\,\sigma_2 \int \frac{ d \theta_{qk} }
 {2 \, \pi } \int_0^\infty \frac{d\omega_k} {\pi} \,
\text{Im} \left[ D^R_{1-loop} (\omega_k,|\theta_{qk}-\theta_{qk'}|)\right]  
 \left[
 \left \lbrace  n_0(\omega_k)+1  \right \rbrace
 f( \omega_{k'} + \omega_{k}, \theta_{qk};\omega_q,\mathbf q)
+  n_0(\omega_k) \,f(\omega_{k'} - \omega_{k}, \theta_{qk};\omega_q,\mathbf q)
 \right] ,
\end{align}
and
\begin{align}
& \frac{\Sigma^> (\omega_{k'}, \mathbf k';\omega_q , \mathbf q)} {e^2\,m }
\nn & = -i\,\sigma_2
  \int \frac{ d\theta_{qk} }
 {2 \, \pi } \int_0^\infty \frac{d\omega_k}{\pi} \,
\text{Im} \left[ D^R_{1-loop}(\omega_k,|\theta_{qk}-\theta_{qk'}|)\right]  
 \Big [
n_0(\omega)\, f(\omega_{k'} + \omega_{k}, \theta_{qk};\omega_q,\mathbf q)
+  \left \lbrace  n_0(\omega_k)+1  \right \rbrace \,
f(\omega_{k'} - \omega_{k}, \theta_{qk};\omega_q,\mathbf q)
\nn & \hspace{ 8.7 cm} -2\, n_0(\omega_k ) -1 \Big]\,. 
\end{align} 
Note that we have assumed that the bosons are always in local thermal equilibrium.
Using the Kramers–Kronig relations and the identity $\mathcal{P}
 \int_0^\infty \frac{d\omega'} {\omega -\omega'} = 0 $,
we get
\begin{align}
& \Sigma^{re}(\omega_{k}, \mathbf k;\omega_q , \mathbf q) 
= - i\,\sigma_2 \int \frac{ d\theta_{qk'} \,d\omega_{k'}} { 2\,\pi^2 }\, 
F_{Landau}( \omega_{k'}-\omega_{k}, \theta_{k'k} )
\, f( \omega_{k'}, \theta_{qk'};\omega_q,\mathbf q)\nn
 \text{ and } & \, 
 \delta \Sigma^{re}(\omega_{k}, \mathbf k;\omega_q , \mathbf q) 
\simeq -i\, \sigma_2 \int \frac{ d\theta_{qk'} \,d\omega_{k'}} { 2\,\pi^2 }\, 
F_{Landau}( \omega_{k'}-\omega_{k}, \theta_{k' k})\,  
\delta f( \omega_{k'}, \theta_{qk'};\omega_q,\mathbf q)  \,,
\end{align}
where 
\begin{align} 
F_{Landau}( \omega_{k}, \theta ) = e^2\,m\,
\text{Re} \Big[ D^R_{1-loop} \big (\omega_{k}, |\theta | \big)
\Big]  
= \frac{ e^2\, m\,{ k_F^4\,\theta }^4 }
{ k_F^6 \,{\theta}^6+\frac{ e^4 \,m^2\,{\omega_{k}}^2 }
{ \pi^2 \,v_F^2 }}  \,.
\end{align}

Integrating over $\omega_k$, and defining $ 
u(\theta_{qk};\nu, \mathbf q) 
= \int_{-\infty}^{\infty} \frac{d\omega_k} {2\,\pi}
\,f(\omega_k, \theta_{qk}; \nu, \mathbf q) $,
the QBE reduces to
\begin{align}
\label{eqqbe}
& \left ( \nu - \sigma_3 \,v_F\, |\mathbf q| \cos \theta_{qk} \right )  \delta u(\theta_{qk};\nu, \mathbf q) 
\nn & = {\rm sgn}(\nu) \int_{-\frac{|\nu|} {2}}^{ \frac{|\nu|} {2}} 
d\omega_k\,\int \frac{  d\theta_{qk'}\,d\omega_{k'} } {4\,\pi^3}  
F_{Landau}( \omega_{k'}-\omega_k,\theta_{k' k})
  \left [
\delta f(\omega_{k'}, \theta_{qk'}; \nu,\mathbf q ) 
- \delta f( \omega_{k'}, \theta_{qk};\nu,\mathbf q)
\right] .
\end{align}
It turns out that
$\delta f(\omega_k, \theta_{qk}; \nu, \mathbf q)$ is
finite only when $|\omega_k| \lesssim |\nu| $ at $T=0$ \cite{kim_qbe}. Therefore, the frequency variable in $ F_{Landau}(\omega_{k}, \theta)$ is cut off by $|\nu| $. In this case, we can introduce a $\nu$-dependent cut-off
$ \theta_{crit} \simeq 
\frac{ e^{2/3} \,m^{1/3}\,|\nu|^{1/3} }
{ \pi^{1/3} \,v_F^{1/3} \,k_F} $ in the angular variable, and approximate $F_{Landau} $ by:
\begin{align}
\frac{ F_{Landau} ( \theta )} {e^2\,m} =\begin{cases}
\frac{ 1 } { k_F^2\, \theta_{crit}^{2} } = 
\frac{\pi^{2/3} \,v_F^{2/3}}  { e^{4/3} \,m^{2/3}\,|\nu|^{2/3} } 
& \text{ for } |\theta| < \theta_{crit}\\
\frac{ 1 } { k_F^2\,\theta^2} & \text{ for } |\theta| > \theta_{crit}
\end{cases}\,,
\end{align}
dropping the frequency-dependence. This simplifies Eq.~\eqref{eqqbe} to
\begin{align}
\label{eqqbe2}
& \left ( \nu - \sigma_3 \,v_F\, |\mathbf q| \cos \theta_{qk} \right )  \delta u(\theta_{qk};\nu, \mathbf q) 
 \simeq \nu 
 \int \frac{  d\theta_{qk'} } { 2\,\pi^2}  \,
F_{Landau}( \theta_{k' k}) \left [
\delta u( \theta_{qk'}; \nu,\mathbf q ) - \delta u(  \theta_{qk};\nu,\mathbf q)
\right] .
\end{align}

We now further decompose the above expression into angular momentum channels
denoted by $\ell$, using the Fourier transforms $ u (\theta;\nu,\mathbf q) 
= \sum \limits_\ell
e^{i\,\ell \, \theta}\, u_\ell ( \nu,\mathbf q) $ [which also implies that
$ \delta f (\omega_k, \theta; \nu,\mathbf q) 
= \sum \limits_\ell e^{i \,\ell \, \theta}
\, \delta f_\ell ( \omega_k; \nu,\mathbf q) $]
and 
$ F_{Landau} (\theta) 
= \sum \limits_\ell
e^{i\,\ell \, \theta}\, F_\ell $.
The QBE thus takes the simplified form:
\begin{align}
 \nu  \,\delta u_\ell ( \nu,\mathbf q)
=  \sigma_3 \,v_F \,|\mathbf q|\,
\frac{ \delta u_{ \ell+1}(\nu, \mathbf q)  + \delta u_{ \ell -1}(\nu, \mathbf q)} 
{2 \left ( 1 + \frac{F_{0 \ell}}{\pi}  \right ) }  \,,\quad
F_{\ell_1 \ell_2} = F_{\ell_1} -F_{\ell_2}\,.
\label{eqqbe3}
\end{align}
Henceforth, we consider the scalar equations obtained from the upper-diagonal components of the above matrix equations.
This infinite set of algebraic equations governs the normal modes of the kinetic theory. These equations are, in fact, analogous to the solutions of the tight-binding models in one dimension.
In order to understand this analogy, we recall that
an electron moving in a one-dimensional lattice, within a tight-binding approximation, is described by the Hamiltonian $H_{TB} = E_0 \sum \limits_n |n\rangle \langle n|
- t_h \sum \limits_n \bigg (\, |n\rangle \langle n+1 | + |n+1 \rangle \langle n|\, \bigg )$, with $t_h$ being the nearest-neighbour hopping amplitude. A general state $|\psi \rangle$ can be expanded in the energy eigenbasis $ \lbrace|n \rangle  \rbrace $ as $|\psi \rangle = \sum \limits_n \psi_n \, |n \rangle$, which leads to an infinite set of coupled linear equations for the coefficients $\lbrace  \psi_n \rbrace $ of the form:
$\left( E-E_0\right) \psi_n = t_h \left( \psi_{n+1}  -\psi_{n-1}\right) $. These kinds of coupled linear equations are solved by an ansatz of the form $\psi_n = e^{i\,k_\lambda n\,a}$, with $a$ being the lattice constant, and $k_\lambda$ denoting the wavenumber. Comparing Eq.~\eqref{eqqbe3} with the tight-binding model, we immediately observe that our QBE can be interpreted as describing a particle hopping in a one-dimensional lattice, with a spatial-dependent hopping amplitude $t_\ell \approx \frac{v_F \,|\mathbf q|} {2 \left ( 1 + \frac{F_{0\ell}}{\pi}  \right ) }$.

The collective modes that might emerge in the collisionless regime, and can propagate like sound modes, were dubbed as ``zero sound'' by Landau for the case of Fermi liquids (as opposed to the first sound that exists in the collision regime). The existence of these modes depends on the interaction parameters $F_{Landau}$, that act as a restoring force. In our case of NFLs, we have managed to find the generalized Landau-interaction parameters, which are well-defined even in the absence of quasiparticles.

\section{Dispersion relations for collective excitations in the colisionless regime}

Let us now study the QBE for longitudinal vibrations with $\ell = 0$.
Due to the presence of $t_\ell$, and observing that $F_{0\ell}$ is a monotonically increasing function of $\ell$,
we now look for modes which correspond to the ``bound
states'' of the analogous tight-binding model, with the energy of the hopping electron being lowest at the lattice point $\ell=0$ (see chapter 13 of Ref.~\cite{feynman3}). 
These correspond to decaying modes with complex $k_\lambda$
(unlike the scattering modes, which are characterized by real values of $k_\lambda$), and give a discrete spectrum. 
Because we want to obtain the dispersion of the zero sound with $\ell = 0$, let us just solve for the modes with $\ell \geq 0$. The analogy with the one-dimensional tight-binding model allows us to make the ansatz
$ \delta u_\ell = e^{-\kappa \left( \ell-1\right)} \,\delta u_1$ for $\ell> 1$ (with $\kappa$ real), and we just need to consider the equations
\begin{align}
\label{eqset3}
\frac{ \nu  \,\delta u_0 } {v_F\,|\mathbf q|}
& = 
\frac{ \delta u_{1} + \delta u_{-1}} 
{2 }  \,,\quad
\frac{\nu  \,\delta u_1 } {v_F\,|\mathbf q|}
= 
\frac{ \delta u_{2}  + \delta u_{ 0}} 
{2 \left ( 1 + \frac{F_{01}}{\pi}  \right ) }  \,,\quad
\frac{  \nu  \,\delta u_2 }
{v_F \, |\mathbf q|} =
\frac{ \delta u_{3}  + \delta u_{1} } 
{2 \left ( 1 + \frac{F_{02} }{\pi}  \right ) } \,,
\end{align}
to solve for the dispersion of zero sound. Furthermore, due to the invariance of $t_\ell$ for $\ell \rightarrow -\ell$, it is reasonable to assume $\delta u_1 = \delta u_{-1}$.
Now, solving these three equations, we get
\begin{align}
\nu = \frac{ \pm  \,\sqrt{2 \,\pi } \,v_F \,|\mathbf q| }
{ \sqrt{\pi + 4 \,F_{01}-2 \,F_{02}  + 
\sqrt{4 \,F_{02}^2
+12 \,\pi \, F_{02}
-8 \,\pi \, F_{01}+\pi ^2} 
}} \,.
\label{soln1}
\end{align}

As long as $ \theta_{crit} \ll 1$, which is usually true for small $e$, $F_{01}, F_{02} \ll 1$ and are just numbers (i.e., independent of $\theta_{crit}$).
Hence, the dispersion of the collective mode has the same behaviour as the zero sound mode (i.e., linear-in-wavenumber dispersion) in a Fermi liquid. Therefore, we conclude that there is no exotic behaviour as far as the zero sound mode is concerned, although we are dealing with an NFL with no quasiparticle excitation.
Let us now try to understand the above result physically.
On the right-hand-side of Eq.~\eqref{eqqbe2}, the term
$\check{t}_1 \equiv
\nu  \int \frac{  d\theta_{qk'} } { 2\,\pi^2}  \,
F_{Landau}( \theta_{k'k})\, \delta u(  \theta_{qk'};\nu,\mathbf q)$ comes from the
Landau-interaction, while the part
$ \check t_2  \equiv
-  \nu \, \delta u(  \theta_{qk};\nu,\mathbf q)
\int \frac{  d\theta_{qk'} } { 2\,\pi^2}  \,
F_{Landau}( \theta_{k'k}) $ corresponds to the contribution
from the real part of the retarded boson self-energy. 
For smooth fluctuations of the generalized Fermi surface displacements $\delta u_\ell$, characterized by $\ell <\ell_{crit} \simeq 1/\theta_{crit}$, we get
$  \check t_1 \propto   |\nu|^{2/3} $.
And it is straightforward to observe that $\check t_2
\propto -|\nu|^{2/3}$. Hence, there is a cancellation between $\check t_1$ and $\check t_2$,
reflected in the fact that $F_{0 \ell} \rightarrow 0$ for $\ell <\ell_{crit}$.
Hence, for the low $\ell$ modes, the QBE reduces to that for a Fermi liquid. For the so-called rough fluctuations $\delta u_\ell \big \vert_{\ell > \ell_{crit}}$, such cancellations will not occur, and the singular behaviour of an NFL is expected to show up in their dispersions.

\section{Effect of weak Coulomb interactions}

The zero sound mode exists only in neutral fermionic fluids. In a
charged system, it is replaced by a plasmon mode.
To study the dispersion of this mode in an NFL, we now add a density-density Coulomb interaction between the electrons, and include its effect within RPA (assuming that the Coulomb interaction is weak). In order to obtain an analytical solution, we follow the formalism described in Ref.~\cite{andy}. Using the arguments in Ref.~\cite{andy}, the Coulomb part is incorporated by modifying Eq.~\eqref{eqqbe3} as
\begin{align}
\left( \nu +i\,\gamma\,\delta_{|\ell|>1} \right) \delta u_\ell ( \nu,\mathbf q)
=  \sigma_3 \,v_F \,|\mathbf q|\,
\frac{ \delta u_{ \ell+1}(\nu, \mathbf q)  + \delta u_{ \ell -1}(\nu, \mathbf q)
+ F_c
\,\delta u_0 \,\delta_{|\ell|,1}
} 
{2 \left ( 1 + \frac{F_0- F_{\ell}}{\pi}  \right ) }  \,,
\quad F_c = \frac{2\,\pi\,\alpha} {\lambda_F \,|\mathbf q|}\,,
\label{eqqbe1}
\end{align}
where $\alpha $ is the effective fine structure constant, and
$\lambda_F  $ is the Fermi wavelength.
Note that we have added a small damping part $\gamma  \ll |\nu| $, which would have originated from the collision integral if we had not set it to zero.
We emphasize that the collisionless regime refers to the scenario when $ |\nu| \gg \gamma$.

The coupled equations now give the solution
\begin{align}
 \nu  &=  
\frac{\pm \,\sqrt{2 \,\pi } \,v_F \,|\mathbf q|  \left( 1 + F_c \right) }
{\sqrt{ \pi + 2 \,\pi \, F_c +2 \left(F_{01} + F_{21} \right) \left( 1 +F_c \right)
+ \sqrt{4 \,F_{02} ^2 \left( 1 +F_c \right)^2
+4 \, \pi \left[
F_0 \left( 1 + 2 \,F_c \right)
-F_2 \left( 3  + 2\, F_c \right)
+2 \,F_1 \right ] 
\left( 1 + F_c \right)
+  \pi^2 \left(2 \, F_c+ 1 \right)^2}}}
\nn & \qquad  -i\, \gamma\frac{ 1+F_c} {\left ( 1 +2 \,F_c \right )^2}\,.
\end{align}
Since we want to see that how the damping of the zero mode is modified by the Coulomb interaction, we have shown the value of the damping part taking into account that
$F_{01}, F_{02} \ll 1$.

In the limit $F_c \ll 1$, the above solution reduces to the expression in Eq.~\eqref{soln1} (plus a small imaginary part $-i\,\gamma$ in the presence of the small contribution $i\,\gamma$ from the collision integral),
behaving as a conventional zero sound mode with linear dispersion.
On the other hand, for $F_c \gg 1$ we get
\begin{align}
& \nu
= \frac{ \pm \, \pi \,v_F \, \sqrt{ |\mathbf q| } \,
\sqrt{\frac{ 2\, \alpha }   {\lambda _F}}}
{\sqrt{ \pi + 2 \,F_{01} -F_{02}  +
\sqrt{\left(F_{02}+\pi \right)^2}  }}
-i\,\gamma\,\frac{ \lambda _F\, |\mathbf q|} {8 \,\pi \, \alpha }\,,
\end{align}
to leading order. This behaves like a plasmon with square-root dispersion,
and its decay rate scales as $|\mathbf q|$ (in agreement with Ref.~\cite{andy}, which deals with a Fermi liquid scenario).
The behaviour found in the two limits again shows that the $\ell=0$ mode is analogous to that in a Fermi liquid.

\section{Summary and conclusion}

In this paper, we have derived the QBE in the collisionless regime, for the NFL system arising at the Ising-nematic QCP. We have used the controlled low-energy field theory developed in Ref.~\cite{Lee-Dalid,ips-uv-ir1}, rather than a large-$N$ expansion (which is an uncontrolled approximation). This has allowed us to find out the dispersion relations of the collective excitations of the Fermi surface, in particular the mode corresponding to zero sound. We have also computed the modification of the dispersion due to the addition of weak Coulomb interactions, which gives the usual long-wavelength plasmon mode. Despite the fact that the fundamental properties of an NFL are strikingly different from those of a normal metal, the zero sound and plasmon modes show the same behaviour as in a Fermi liquid. The generalized Landau-interaction parameters lead only to minor changes
in the speed of zero sound / plasmon, while the power-law-dependence on momentum remains the same.
 
We would like to emphasize that we have performed our computations right at the quantum phase transition point, which takes place at $T=0$. However, the calculations can be extended to the $T>0$ regime, in the same spirit as done in the appendix of Ref.~\cite{kim_qbe}, where the quantum critical region extends in a fan-shaped area, which shows strange metal behaviour through strange dependence of transport properties (e.g. resistance) on temperature. Of course, if we move away from this quantum critical region, we will see a crossover to Fermi liquid behaviour. But as far as the characteristics of the zero sound and plasmon modes are
concerned, we expect to observe the same behaviour in all these regions, as we have argued that the numbers $F_{01}$ and $F_{02}$ are too small to warrant any significant change in the velocities of these collective modes. In other words, a deviation from the quantum critical region is not expected to be reflected in the behaviour of these collective modes of the Fermi surface.

Although sound modes have been observed in liquid $^3$He \cite{abel} and strongly interacting cold atomic gases \cite{eva}, it might be challenging to carry out similar experiments on the chemically complex high-$T_c$ materials, where NFL phases like the Ising-nematic QCP are expected to emerge. In Ref.~\cite{andy}, the authors have presented a plausible set-up for an experiment, which could potentially detect the key qualitative signatures of sound modes, in a strongly interacting charge-neutral electron-hole plasma (e.g. Dirac fluid in graphene). Such systems have sharply-defined sound modes, and can be observed upon injecting energy into the system at a modulated rate. They have also showed that at the charge neutrality point, the acoustic resonances are essentially immune to any possible long-ranged Coulomb screening, and are also robust to moderate amounts of disorder. One can try to design similar set-ups for the Ising-nematic quantum critical region (studied in this paper), using high-T$_c$ materials, which would be amenable to host the NFL phases. However, it is not {\it a priori} guaranteed that what is expected for a cleaner and simpler system like graphene, would still be applicable for a complex high-T$_c$ material. This is especially because the properties of the latter cannot be tuned {\it in situ} to analyze various phases as a function of the tuning parameter $r_c$ (which can be doping, magnetic field, pressure and so on). A possible way out could be to consider similar phases which are expected to emerge in more tunable systems like moir\'e heterostructures (e.g. magic angle twisted bilayer graphene \cite{cao1,cao2,ips-tbg}).

In future, it will be interesting to derive the $T>0$ results for the QBE, and also look at the collision regime (when the collision integral cannot be neglected).
Another direction is to incorporate impurity-scattering, and derive its effects on the various collective modes, because the collision integral plays an important role when we include impurities. A challenging plan is to incorporate the invariant measure approach (IMA) \cite{ips-klaus} to find such behaviour in the presence of disorder, which has so far not been formulated for NFLs.
Last but not the least, it will be worthwhile to study the dynamics of the plasmon modes in NFLs which arise at a band-touching point \cite{polini,ips_qbt_plasmons,krempa,Herbut-PRB} (Fermi point), using our QBE approach.

\section{Acknowledgments}
We thank Andrew Lucas and Klaus Ziegler for useful comments.

\appendix

\bibliography{biblio}

\end{document}